\documentstyle[seceq,epsf,wrapft,twoside]{ptptex}
\notypesetlogo  
\title{Chiral Symmetry  Restoration and the $\sigma$-meson
\footnote{
Invited talk at the International Workshop on 
 ``Possible Existence of the Light $\sigma$ Resonance and 
its Implication to Hadron Physics'', (June 11-14, 2000,
  Yukawa Institute for Theoretical Physics, Kyoto, Japan).}
}
\author{%
Tetsuo {\sc Hatsuda}$^{\dagger}$ and Teiji {\sc Kunihiro}$^{\dagger\dagger}$}
\inst{%
$^{\dagger}$ Department of Physics, Univ. of Tokyo,\\ Tokyo 113-0033, Japan\\
$^{\dagger\dagger}$ Yukawa Institute for Theoretical Physics, Kyoto Univ.,
\\ Kyoto 606-8502, Japan }
\abst{The $\sigma$-meson, which is at best a broad resonance in the 
 vacuum, may suffer a substantial red-shift and appear as a 
  soft and narrow collective mode at finite temperature and baryon densities.
  The physics behind this softening phenomena and its relation to the
   partial restoration of chiral symmetry are discussed.
    Possible experiments in laboratory to test the idea are also
    summarized.}

\begin{document}
\maketitle

\setcounter{tocdepth}{4}

\section{Introduction}

One of the most intriguing phenomena in QCD is the 
dynamical breaking of chiral  symmetry (DB$\chi$).
This explains the existence of the pion and dictates
most of the low energy phenomena in hadron physics.
   DB$\chi$ is associated with the condensation of quark - anti-quark
pairs in the vacuum, 
 $\langle \bar{q}q \rangle$, which is analogous to the condensation of 
 Cooper pairs 
 in the theory of  superconductivity \cite{NJL}.
 As the temperature ($T$) and/or the baryon density increase,
 the QCD vacuum undergoes a phase transition to the chirally
 symmetric phase where $\langle \bar{q}q \rangle$ vanishes \cite{HK94}.
 For massless 2-flavors at finite $T$ with vanishing baryon density,
  the chiral transition is likely to be of  
   second order from the renormalization group analysis 
   with the universality hypothesis \cite{PW} and from the 
   direct lattice QCD simulations \cite{finiteT}. The chiral transition at  
   zero $T$ with finite baryon density is not well understood,
   but some of the effective theories
   suggest the first order transition at several times the nuclear matter
    density.

The  general wisdom of many-body physics \cite{GK}  tells us that 
the fluctuation of the order parameter 
  becomes large as
 the system approaches to the critical point of a second order 
 or weak first order phase transition.
 In QCD,  the fluctuations of 
  the phase and the amplitude of the chiral order parameter 
 $\langle \bar{q}q \rangle $
  correspond to  the pion ($\pi$) and the sigma-meson ($\sigma$),
   respectively.
 Their vital roles in the 
  {\em dynamical} phenomena near the critical point of
   chiral phase transition at finite $T$ 
  was first studied  by the present authors in ref. \cite{HK85};
 it was shown in \cite{HK85} that  the chiral restoration gives rise to a 
  softening (the red-shift) of the $\sigma$, which in turn  leads to
   a small  $\sigma$-width 
  due to the suppression of the phase-space of the decay
  $\sigma \rightarrow 2 \pi$. Therefore, the $\sigma$ may 
  appear as a narrow resonance at finite $T$
  although it is at best 
  a very broad resonance in the free space with a  width comparable to its
  mass \cite{pipi,comment}.
   For the phenomenological applications of the above  
 idea (the softening and narrowing of $\sigma$  at finite $T$) in relation 
 to the relativistic heavy ion collisions, see  \cite{later}.

\begin{figure}[t]
  \epsfxsize=10 cm
 \centerline{\epsffile{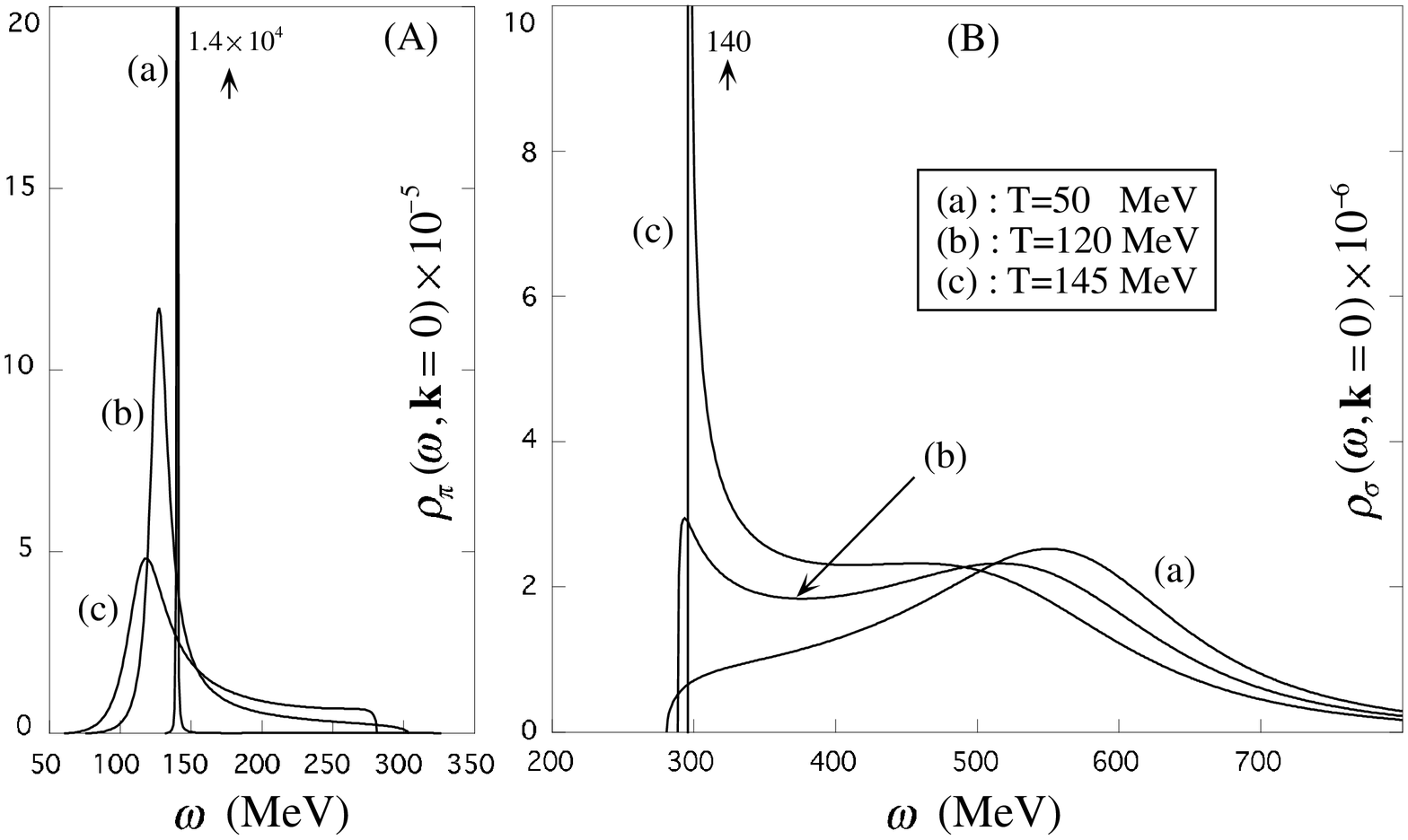}}
 \caption{ Spectral functions in the $\pi$ channel (A) and in the
 $\sigma$ channel (B) for $T$=50, 120,  and 145 MeV. \cite{CH98}
 $\rho_{\sigma}$ in (B) shows only a broad bump at low $T$ (a), while
 the spectral concentration is developed as $T$ increases as
  (a)$\rightarrow$(b)$\rightarrow$(c).}
  \label{fig:1}
\end{figure}

 Further theoretical analysis \cite{CH98}
  by taking  account the
  coupling $\sigma \leftrightarrow 2\pi$  shows that  
 (i) the spectral function  is the most relevant quantity
 for studying the nature of $\sigma$ (which was already discussed in 
 \cite{HK85}), and (ii)  the spectral function of  $\sigma$ 
 has a characteristic enhancement just above the
  two-pion threshold near and below the critical temperature $T_c$  of 
   chiral transition. This enhancement
     may be measured by the di-photon
    spectrum from the hot plasma created in the 
     relativistic heavy ion collisions \cite{CH98,VOL}.
     In Fig.\ref{fig:1}, shown are the 
     spectral functions in the $\pi$- and the $\sigma$-channels at finite $T$
 calculated in the $O(4)$ linear $\sigma$ model:
 The apparent two characteristic features are 
 the broadening of the pion peak (Fig.1(A)) and the
    spectral concentration just above the 2$\pi$ threshold 
 ($\omega \simeq  2m_{\pi}$)  in the $\sigma$-channel (Fig.1(B)).
   
 The notion of the chiral soft-mode
    has recently  been applied to finite baryon density in \cite{HKS}.
 It was shown  that the near-threshold enhancement could be seen as long as 
 the partial restoration of chiral symmetry occurs,
 irrespective of the order of the phase transition.
   Since $\langle \bar{q}q \rangle$ is expected to decrease almost by 35\%
  in nuclear matter as shown below,
    one may observe the near-threshold
     enhancement in the 
  di-pion and di-photon spectra in hadron-nucleus and photon-nucleus
   reactions as was pointed out in  \cite{kuni95}.
   
 In the following, we will focus on the recent development
 at finite baryon density and illustrate the essential ideas behind
 the physics of in-medium $\sigma$.

\section{In-medium spectral function of $\sigma$}

\subsection{Quark condensate in nuclear matter}

 $\langle \bar{q}q \rangle$
 at finite baryon density ($\rho$) obeys  an exact theorem in QCD  \cite{DL}:
 \begin{eqnarray}
 \label{cond-rho}
{\langle \bar{q} q \rangle \over
 \langle \bar{q} q \rangle_{0} }
= 1 - {\rho \over f_{\pi}^2 m_{\pi}^2 } \left[ \Sigma_{\pi N}
 + m {d \over dm} \left( {E(\rho)  \over A } \right) \right] ,
 \end{eqnarray}
where $\Sigma_{\pi N}= 45 \pm 10 $ MeV is the pion-nucleon sigma term
 and $E(\rho)/A$ is the nuclear binding energy per particle with $m$
being the current quark mass. 
$\langle \bar{q} q \rangle_{0}\simeq -$(225 MeV)$^3$
 denotes the chiral condensate in the vacuum.
The  density-expansion of the right hand side of (\ref{cond-rho})
 gives a reduction of almost 35 \% of $\langle \bar{q} q \rangle $
 already at the nuclear matter density $\rho_0 = 0.17
 $fm$^{-3}$; notice that the first density modification comes linearly
in $\rho$.

\subsection{Basic idea of the spectral enhancement}

A direct evidence of partial chiral restoration could be an
 enhancement of the spectral strength in the scalar-isoscalar 
channel near the 2$\pi$ threshold as shown in Fig.1(B).
 The physics behind this  phenomena is the following:
  Suppose we have 
 Landau free-energy of a double-well type  written in terms of the 
 order parameter $\sigma \sim \langle \bar{q}q \rangle $,
\begin{eqnarray}
V (\sigma) = - {a \over 2} \sigma^2 + {b \over 4} \sigma^4 ,
  \end{eqnarray}
where $a$ is positive in the vacuum but
 changes the sign at the critical point, while $b$ remains
positive.
 The minimum of the effective potential $\sigma_0$  and the curvature at the
 minimum read
\begin{eqnarray}
\sigma_0 = \sqrt {a \over b} , \ \ \ \ 
{1 \over 2} \left. {d^2 V(\sigma) \over d\sigma^2 }\right|_{\sigma = \sigma_0}
 = a \ \ .
  \end{eqnarray}
Therefore, as $a$ becomes small, not only the order parameter
 $\sigma_0$ but also the curvature  decreases.
 This is nothing but the softening of the oscillational mode 
of the order parameter
 associated with the second order phase transition \cite{HK94}.

 In the real world, the situation is not that simple,
 since the $\sigma$ has a large width from the decay $\sigma \rightarrow 2 \pi$.
 Nevertheless, there is an interesting possibility that
 the spectral function just above the 2$\pi$ threshold could
 be enhanced due to the change of $\sigma_0$.
 To describe the general
 features of this spectral enhancement, let us 
 consider the propagator 
 of the scalar-isoscalar $\sigma$-meson at rest in the medium:
$D^{-1}_{\sigma} (\omega)= \omega^2 - m_{\sigma}^2 - $
$\Sigma_{\sigma}(\omega;\rho),$
where $m_{\sigma}$ is the mass of the $\sigma$ in the tree-level, and
$\Sigma_{\sigma}(\omega;\rho)$ is 
the loop corrections
 in the vacuum as well as in the medium.
 The corresponding spectral function is given by 
$\rho_{\sigma}(\omega) = - {1 \over \pi } {\rm Im} D_{\sigma}(\omega).$
Near the 2$\pi$ threshold, the imaginary part 
 in the one-loop order	 reads 
\begin{eqnarray}
{\rm Im} \Sigma_{\sigma} \propto \theta(\omega - 2 m_{\pi}) \ 
	 \sqrt{1 - {4m_{\pi}^2 \over \omega^2}} .
\end{eqnarray}

 When chiral symmetry is being restored,
 $m_{\sigma}^*$ (``effective mass'' of $\sigma$ 
 defined as a zero of the real part of the propagator
 ${\rm Re}D_{\sigma}^{-1}(\omega = m_{\sigma}^*)=0$)
  approaches to $ m_{\pi}$.  Therefore,
 there exists a density $\rho_c$ at which 
 ${\rm Re} D_{\sigma}^{-1}(\omega = 2m_{\pi})$
 vanishes even before the complete $\sigma$-$\pi$
 degeneracy is realized; 
$ {\rm Re} D_{\sigma}^{-1} (\omega = 2 m_{\pi}) =
 [\omega^2 - m_{ \sigma}^2 -
 {\rm Re} \Sigma_{\sigma} ]_{\omega = 2 m_{\pi}} = 0.$
At this point, the spectral function can be  solely represented by the
 imaginary part of the self-energy;
\begin{eqnarray}
\rho_{\sigma} (\omega \simeq  2 m_{\pi}) 
 =  - {1 \over \pi \ {\rm Im}\Sigma_{\sigma} }
 \propto {\theta(\omega - 2 m_{\pi}) 
 \over \sqrt{1-{4m_{\pi}^2 \over \omega^2}}},
\end{eqnarray}
which shows an enhancement of the spectral function at the $2\pi$
threshold. One should note that this enhancement is 
owing to the  partial restoration of chiral symmetry and hence generic.

\subsection{A calculation based on the O(4)  $\sigma$-model}

To make the argument more quantitative,
let us evaluate $\rho_{\sigma}(\omega)$ in 
 the O(4) linear $\sigma$-model:
\begin{eqnarray}
\label{model-l}
{\cal L}  =  {1 \over 4} {\rm tr} [\partial M \partial M^{\dagger}
 - \mu^2 M M^{\dagger} 
  - {2 \lambda \over 4! } (M M^{\dagger})^2   - h (M+M^{\dagger}) ],
\end{eqnarray}
where tr is for the flavor index and  
$M = \sigma + i \vec{\tau}\cdot \vec{\pi}$.
 Although the model has only a limited number of parameters
 and is not a precise low energy representation of QCD,  
 it can  describe the
 pion dynamics qualitatively well up to 1GeV  \cite{BW}.
 The coupling constants $\mu^2, \lambda$ and $h$ have
 been determined in the
 vacuum to reproduce $f_{\pi}=93$ MeV, $m_{\pi}=140$ MeV as well as
 the s-wave $\pi$-$\pi$ scattering phase shift in the one-loop order.
 We parameterize the chiral condensate in nuclear matter
 $\langle \sigma \rangle$ as
\begin{eqnarray}
\langle \sigma \rangle \equiv  \sigma_0 \ \Phi(\rho).
\end{eqnarray}
Depending on how one relates $\langle \sigma \rangle$ with
 $\langle \bar{q}q \rangle$,  $\Phi(\rho = \rho_0)$ may take a value
in the range  0.6 $\sim$ 0.9 \cite{HK94}.

 \begin{figure}[b]
 \epsfxsize=7 cm
   \centerline{\epsffile{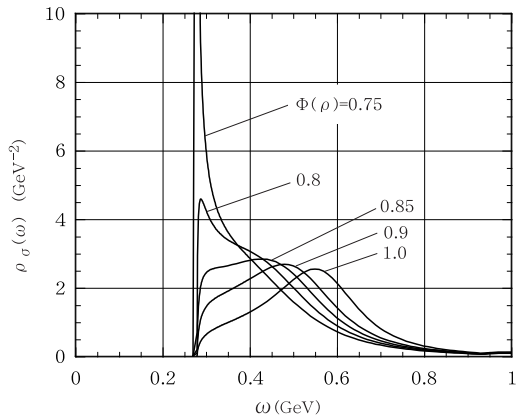}}
 \caption{Spectral function for $\sigma$ and  the 
 real part of the inverse propagator for several values of
 $\Phi = \langle \sigma \rangle / \sigma_0$. \cite{HKS}
 Similar spectral concentration with Fig.1(B) can be seen
 as the baryon density increases.
}
\label{fig:2}
\end{figure}

The in-medium spectral function in the $\sigma$-channel 
 calculated in eq.(\ref{model-l})  in the one-loop level are shown 
  in Fig.2: The characteristic enhancement of the spectral
 function is seen just above the 2$\pi$ threshold as expected.
 To confirm this threshold enhancement experimentally,
 measuring 2$\pi^0$ and 
$2\gamma$ in experiments with hadron/photon beams off
 the  heavy nuclear targets are useful \cite{kuni95}. 
 Measuring $\sigma \rightarrow 2 \pi^0 \rightarrow
  4\gamma$ is experimentally feasible 
 \cite{4gamma}, which is free from the $\rho$ meson background
  inherent in the $\pi^+\pi^-$ measurement.
 Measuring $\sigma \rightarrow 2 \gamma$
  is also interesting because of the small final state
 interactions, although the branching ratio is small.
 (One needs also to fight with the large 
 background of photons mainly coming from $\pi^0$s.)
 Nevertheless,  if the enhancement is prominent,
 there is a chance to find the signal.  
  There is also a possibility that one can detect dilepton 
 through the scalar-vector mixing in matter: $\sigma \to \gamma^* \to
 e^+ e^-$ \cite{kuni95}. In this case,
  the dileptons are produced  only when
 $\sigma$ has a finite three momentum.
  Another possible experiment is the formation of the
  $\sigma$ mesic nuclei through the 
   the nuclear reactions such as the (d, t)  reaction.
  The incident kinetic energies of the
  deuteron in the laboratory
 system $E$ can be estimated to be
  $1.1 {\rm GeV} < E < 10$ GeV, 
  to cover the spectral function 
 in the range  $2m_{\pi} < \omega < 750$ MeV \cite{HKS,hiren}.

 \begin{figure}[b]
 \epsfxsize=8 cm 
  \centerline{\epsffile{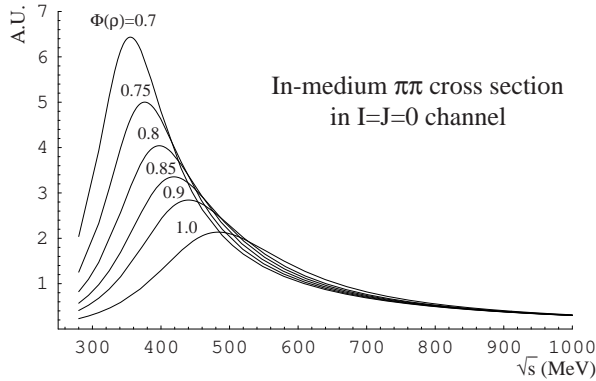}}
 \caption{ In-medium $\pi\pi$ cross section in the $I=J=0$ channel
 for different values of $\Phi(\rho)$.
    The cross section is shown in the arbitrary unit (A.U.).\cite{JHK} 
}
\label{fig:3}\end{figure}

Recently  CHAOS collaboration  \cite{CHAOS} reported the data on the
$\pi^{+}\pi^{\pm}$
invariant mass distribution $M^A_{\pi^{+}\pi^{\pm}}$ in the
 reaction $A(\pi^+, \pi^{+}\pi^{\pm})A'$ with the 
 mass number $A$ ranging
 from 2 to 208: They observed that
the   yield for  $M^A_{\pi^{+}\pi^{-}}$ 
 near the 2$\pi$ threshold is close to zero 
for $A=2$, but increases dramatically with increasing $A$. They
identified that the $\pi^{+}\pi^{-}$ pairs in this range of
 $M^A_{\pi^{+}\pi^{-}}$ is in the $I=J=0$ state.
This experiment was
 originally motivated by a possibility of strong
 $\pi\pi$ correlations in nuclear matter \cite{GSI}.
  However,  the state-of-the-art
  calculations using the nonlinear chiral
  lagrangian together with  $\pi N$ many-body
  dynamics do not reproduce the cross sections in the $I=0$ and
  $I=2$ channels simultaneously  \cite{WO};
  the final state interactions of the emitted two pions in nuclei
  give rise to a slight enhancement of the cross section in the
  $I=0$ channel, but is
  not sufficient to reproduce the experimental  data.
 This indicates that
 some additional mechanism such as the partial restoration of chiral
 symmetry first proposed in \cite{HKS} 
  may be
 relevant for explaining the data (see also the later studies \cite{SHUCK}).

In a recent paper, to make a close connection between the 
 idea of the spectral enhancement and the CHAOS data,
  the in-medium $\pi$-$\pi$ cross section has been calculated in 
   the linear and non-linear $\sigma$ models \cite{JHK}.
   It was shown that, in {\em both} cases, substantial enhancement of the 
   pion-pion correlation in the $I=J=0$ channel near the threshold can be seen 
   due to the partial restoration of chiral symmetry, and 
   an effective 4$\pi$-N-N vertex responsible for the enhancement
    is identified. In Fig. 3, the in-medium $\pi$-$\pi$ cross section
   in the $\sigma$-channel
  calculated in the O(4) linear $\sigma$ model is shown, where
     the same parameter sets are used with Fig.1 and Fig.2.
  It is 
  of great importance to make
 an extensive analysis of the $\pi \pi$ interaction in nuclear matter
 with  the new vertex which has not been considered before in the
  analysis of the CHAOS data.

\section{Conclusion}
 
 The light $\sigma$-meson  does not show up 
 clearly in the free space because of its large  width due to 
the strong coupling with two pions. However,
 it may appear as a soft and narrow
 collective mode in the hadronic medium 
when the chiral symmetry is (partially) restored.
 A characteristic signal is the enhancement of the spectral function
 and the $\pi$-$\pi$ cross section near the  $2\pi$ threshold. They
 could be observed in the hadronic
 reactions with heavy nuclear targets as well as in the heavy
 ion collisions.  Detecting such signal provides us with a 
 better understanding of the non-perturbative structure of the 
 QCD vacuum and its quantum fluctuations.

\vspace{0.2cm}
This work is supported by the Grants-in-Aids of the Japanese
Ministry of Education, Science and Culture (No. 12640263 and 
12640296).

\end{document}